# Coordinate space methods for the evaluation of Feynman diagrams in lattice field theories

Martin Lüscher

Deutsches Elektronen-Synchrotron DESY
Notkestrasse 85, D-22603 Hamburg, Germany

Peter Weisz

Max-Planck-Institut für Physik
Föhringer Ring 6, D-80805 München, Germany

**Abstract**

We describe an efficient position space technique to calculate lattice Feynman integrals in infinite volume. The method applies to diagrams with massless propagators. For illustration a set of two-loop integrals is worked out explicitly. An important ingredient is an observation of Vohwinkel that the free lattice propagator can be evaluated recursively and is expressible as a linear function of its values near the origin.

# 1. Introduction

Perturbation theory plays an important conceptual and practical rôle in lattice field theory. The continuum limit can be studied in this framework and numerical simulations must often be supplemented by perturbative calculations to obtain the desired physical result. The evaluation of lattice Feynman diagrams is not easy in general, because the Feynman integrands tend to be rather complicated functions of the loop and external momenta. Standard tools, such as the Feynman parameter representation and partial integration methods, are usually not very helpful on the lattice.

In a recent computation of the relation between the bare lattice coupling and the $\overline{\rm MS}$ coupling in $SU(N)$ gauge theories [1] we have made extensive use of some new position space techniques. They are expected to be more generally applicable and so it is our aim in the present paper to describe these methods in some detail. Basically we discuss a number of examples and leave it to the reader to adapt the methods to the case of interest. In other words, we do not attempt to set up a general strategy for the computation of lattice Feynman diagrams. We feel that more experience must be accumulated before this is possible.

Throughout this paper we assume that the theory considered lives on a four-dimensional hypercubic lattice $\Lambda$ with spacing $a = 1$. We only discuss diagrams with free propagators equal to $1/\hat{p}^2$, where $p = (p_0, \ldots, p_3)$ denotes the line momentum and

$$\hat{p}_\mu = 2\sin\left(\tfrac{1}{2}p_\mu\right), \qquad \hat{p}^2 = \sum_{\mu=0}^{3} \hat{p}_\mu^2. \qquad (1.1)$$

This covers the case of (pure) lattice gauge theories with standard Wilson action. Fermion propagators are different and some additional work will be required before our methods can be applied to lattice QCD.

The basic ideas to be discussed in this paper are best explained by considering the simple two-loop integral

$$A_0 = \int_{-\pi}^{\pi} \frac{{\rm d}^4k}{(2\pi)^4} \frac{{\rm d}^4q}{(2\pi)^4} \frac{1}{\hat{k}^2\hat{q}^2\hat{r}^2}, \qquad r = -k - q. \qquad (1.2)$$

$A_0$ can be calculated directly using a Monte Carlo program, for example, but it would be difficult to obtain the integral very accurately in this way. Now if we introduce the position space propagator

$$G(x) = \int_{-\pi}^{\pi} \frac{\mathrm{d}^4 p}{(2\pi)^4} \frac{\mathrm{e}^{ipx}}{\hat{p}^2}, \qquad (1.3)$$

the integral may be rewritten in the form

$$A_0 = \sum_{x \in \Lambda} G(x)^3. \qquad (1.4)$$

Since $G(x)$ falls off like $1/x^2$ at large $x$, the sum is absolutely convergent and a first approximation to $A_0$ may be obtained by summing over all points $x$ with say $|x| \leq 20$. Taking the detailed asymptotic form of $G(x)$ into account, one can then systematically improve on the accuracy until the desired level of precision is reached.

For this procedure to work out one must be able to calculate the propagator $G(x)$ accurately at all lattice points $x$ that are not too far away from the origin. An elegant solution to this problem has been given by Vohwinkel [2]. His method allows us to obtain the propagator to essentially any desired precision with a minimal amount of computer time.

In sect. 2 we discuss the free propagator in detail and explain Vohwinkel's method. We then address the question of how to calculate lattice sums accurately and consider various examples to illustrate the procedure (sect. 3). The position space method is also useful to calculate Feynman diagrams with non-zero external momenta, particularly in the continuum limit. This is demonstrated in a number of cases in sect. 4. Some concluding remarks are collected in sect. 5. Much of the technical material has been deferred to a series of appendices to make the paper more readable.

## 2. The free propagator

We begin by discussing the behaviour of $G(x)$ at large $x$ and then derive a recursion relation which allows us to express the propagator through its values close to the origin. After some further steps a strategy for the numerical calculation of $G(x)$ emerges which proves to be very efficient and simple to program.

## 2.1 Asymptotic expansion at large $x$

In terms of the forward and backward lattice derivatives,

$$\nabla_\mu f(x) = f(x + \hat{\mu}) - f(x), \qquad \nabla^*_\mu f(x) = f(x) - f(x - \hat{\mu}) \qquad (2.1)$$

(where $\hat{\mu}$ denotes the unit vector in direction $\mu$), the lattice laplacian is given by

$$\Delta = \sum_{\mu=0}^{3} \nabla^*_\mu \nabla_\mu. \qquad (2.2)$$

$G(x)$ is a Green function for $\Delta$, i.e. it satisfies

$$-\Delta G(x) = \begin{cases} 1 & \text{if } x = 0, \\ 0 & \text{otherwise.} \end{cases} \qquad (2.3)$$

In the continuum limit (which here amounts to taking $x$ to infinity) we thus expect that $G(x)$ converges to $(4\pi^2 x^2)^{-1}$, the Green function of the continuum laplacian on $\mathbb{R}^4$.

A rigorous derivation of the large $x$ behaviour of $G(x)$ confirms this (appendix A). One may also work out the subleading terms and obtains the asymptotic expansion

$$G(x) \underset{x \to \infty}{\sim} \frac{1}{4\pi^2 x^2} \left\{ 1 - \frac{1}{x^2} + 2\frac{x^4}{(x^2)^3} \right.$$

$$\left. - 4\frac{1}{(x^2)^2} + 16\frac{x^4}{(x^2)^4} - 48\frac{x^6}{(x^2)^5} + 40\frac{(x^4)^2}{(x^2)^6} + \ldots \right\}, \qquad (2.4)$$

which includes all terms up to order $|x|^{-6}$. For notational convenience the shorthand

$$x^n = \sum_{\mu=0}^{3} (x_\mu)^n \qquad (2.5)$$

has been used in eq.(2.4).

*2.2 Recursion relation*

The key to Vohwinkel's method is the observation that

$$(\nabla_\mu^* + \nabla_\mu)G(x) = x_\mu H(x), \qquad (2.6)$$

where

$$H(x) = \int_{-\pi}^{\pi} \frac{\mathrm{d}^4 p}{(2\pi)^4} e^{ipx} \ln(\hat{p}^2) \qquad (2.7)$$

is independent of $\mu$. Eq.(2.6) thus represents a set of 4 relations for $G$ and $H$ of which one may be used to eliminate $H$. For example, by summing over $\mu$ and using eq.(2.3) one obtains

$$H(x) = (2/\rho) \sum_{\mu=0}^{3} [G(x) - G(x - \hat{\mu})], \qquad \rho = \sum_{\mu=0}^{3} x_\mu, \qquad (2.8)$$

and hence

$$G(x + \hat{\mu}) = G(x - \hat{\mu}) + (2x_\mu/\rho) \sum_{\nu=0}^{3} [G(x) - G(x - \hat{\nu})]. \qquad (2.9)$$

Of course this relation is only valid at the lattice points $x$ where $\rho$ does not vanish.

The propagator $G(x)$ is independent of the sign and the order of the coordinates $(x_0, \ldots, x_3)$. We can hence restrict attention to the points $x$ with $x_0 \geq x_1 \geq x_2 \geq x_3 \geq 0$. In this sector eq.(2.9) is a recursion relation which allows us to express $G(x)$ as a linear combination of

$$G(0,0,0,0), G(1,0,0,0), \ldots, G(1,1,1,1), \qquad (2.10)$$

i.e. by the values of the propagator at the corners of the unit hypercube.

The recursion alone does not imply any restrictions on these 5 initial values. When further properties of the propagator are taken into account one may however show that

$$G(0,0,0,0) - G(1,0,0,0) = 1/8, \qquad (2.11)$$

$$G(0,0,0,0) - 3G(1,1,0,0) - 2G(1,1,1,0) = 1/\pi^2, \qquad (2.12)$$

$$G(0,0,0,0) - 6G(1,1,0,0) - 8G(1,1,1,0) - 3G(1,1,1,1) = 0. \qquad (2.13)$$

The first of these relations is equivalent to eq.(2.3) at $x = 0$. An amusing way to deduce the other two is given in appendix B. Thus, from the 5 initial values 3 can be eliminated and the propagator is obtained in the form

$$G(x) = r_1(x)G(0,0,0,0) + r_2(x)G(1,1,0,0) + r_3(x)/\pi^2 + r_4(x), \qquad (2.14)$$

where $r_1(x), \ldots, r_4(x)$ are rational numbers recursively computable through eq.(2.9).

*2.3 Numerical computation of $G(x)$*

Using MAPLE or some other algebraic programming language, it is straightforward to write a program which calculates the coefficients $r_1(x), \ldots, r_4(x)$ exactly. We are then left with the task to compute $G(0,0,0,0)$ and $G(1,1,0,0)$ numerically. It is necessary to obtain these numbers very accurately, since the significance losses in eq.(2.14) can be huge. The reason for this is that $r_1(x), \ldots, r_4(x)$ are exponentially increasing at large $x$ while $G(x)$ remains bounded.

An interesting observation now is that we can turn the tables and use the rapid growth of the coefficients $r_1(x), \ldots, r_4(x)$ to calculate $G(0,0,0,0)$ and $G(1,1,0,0)$ in a most efficient way. To this end we consider the lattice points $x = (n,0,0,0)$ and $y = (n,1,0,0)$, where $n$ is an adjustable integer. Inspection shows that the associated coefficients $r_k(x)$ and $r_k(y)$ are roughly proportional to $10^n$. So if we express $G(0,0,0,0)$ and $G(1,1,0,0)$ as linear combinations of $G(x)$, $G(y)$, $1/\pi^2$ and 1 [by solving eq.(2.14) and the corresponding equation for $G(y)$], the coefficients multiplying $G(x)$ and $G(y)$ will be of order $10^{-n}$. Neglecting these terms one thus obtains $G(0,0,0,0)$ and $G(1,1,0,0)$ to this precision.

Note that this method of calculating $G(0,0,0,0)$ and $G(1,1,0,0)$ is exponentially convergent. Moreover the error can be estimated rigorously since the neglected terms are known and can be bounded easily. With a negligible amount of computer time one thus obtains

$$G(0,0,0,0) = 0.154933390231060214084837208107\ldots, \qquad (2.15)$$

$$G(1,1,0,0) = 0.012714703770934215428228758391\ldots, \qquad (2.16)$$

and it would cost little more time to get the next 30 digits of these numbers.

## 3. Evaluation of lattice sums

We now discuss how to calculate lattice sums such as the one appearing in eq.(1.4). As a preparation we first consider the zeta functions $Z(s,h)$ associated with the hypercubic lattice $\Lambda$.

*3.1 Zeta functions*

Let $h(x)$ be a harmonic homogeneous polynomial in the real variables $x = (x_0, \ldots, x_3)$ with even degree $d$. Examples of such polynomials are

$$h_0(x) = 1, \tag{3.1}$$

$$h_1(x) = 2x^4 - (x^2)^2, \tag{3.2}$$

$$h_2(x) = 16x^6 - 20x^2 x^4 + 5(x^2)^3, \tag{3.3}$$

$$h_3(x) = 560(x^4)^2 - 560x^2 x^6 + 60(x^2)^2 x^4 - 9(x^2)^4 \tag{3.4}$$

[cf. eq.(2.5)]. For complex $s$ with $\operatorname{Re} s > 2 + d/2$ we define a generalized zeta function through

$$Z(s,h) = {\sum_{x \in \Lambda}}' h(x)(x^2)^{-s}, \tag{3.5}$$

where the primed summation symbol implies that the point $x = 0$ should be omitted. Note that the sum (3.5) is absolutely convergent in the specified range of $s$.

The zeta functions can be calculated as follows. We first introduce the heat kernel

$$k(t,h) = \sum_{x \in \Lambda} h(x) e^{-\pi t x^2} \tag{3.6}$$

and rewrite eq.(3.5) in the form

$$Z(s,h) = \frac{\pi^s}{\Gamma(s)} \int_0^\infty dt\, t^{s-1} \left[ k(t,h) - h(0) \right]. \tag{3.7}$$

Using the Poisson summation formula

$$\sum_{x \in \Lambda} e^{-iqx} e^{-\pi t x^2} = t^{-2} \sum_{x \in \Lambda} e^{-(q+2\pi x)^2/(4\pi t)} \tag{3.8}$$

and the property that $h$ is harmonic, we deduce that the heat kernel obeys

$$k(t,h) = (-1)^{d/2} t^{-d-2} k(1/t, h). \tag{3.9}$$

Now if we split the integral (3.7) in two parts and apply the identity (3.9), the representation

$$Z(s,h) = \frac{\pi^s}{\Gamma(s)} \left\{ \frac{2h(0)}{s(s-2)} \right.$$

$$\left. + \int_1^\infty dt \, [t^{s-1} + (-1)^{d/2} t^{d-s+1}][k(t,h) - h(0)] \right\} \tag{3.10}$$

is obtained. In the integration range $1 \leq t < \infty$ the series (3.6) is rapidly convergent. Eq.(3.10) is hence suitable for numerical evaluation. In the cases of interest the zeta functions can in this way be calculated to (say) 16 significant digits with little effort.

*3.2 Computation of $A_0$*

Since the propagator $G(x)$ is known accurately, we can obtain an estimate of the sum (1.4) by summing all terms with $|x| \leq 20$. The result of this calculation is accurate to a relative precision of about $10^{-4}$, i.e. the approximation is already quite good. One could of course do better by including more terms, but since the sum is only quadratically convergent this is not the most efficient way to proceed.

The convergence can be accelerated by making use of the known asymptotic form of the propagator. From eq.(2.4) we deduce that

$$G(x)^3 \underset{x \to \infty}{=} \{G(x)^3\}_{\text{as}} + O(|x|^{-12}), \tag{3.11}$$

where

$$\{G(x)^3\}_{\text{as}} = \frac{1}{(4\pi^2)^3} \left\{ \left[ \frac{1}{(x^2)^3} + \frac{3}{10(x^2)^5} \right] h_0(x) + \left[ \frac{3}{(x^2)^6} + \frac{24}{7(x^2)^7} \right] h_1(x) \right.$$

$$\left. - \frac{3}{4(x^2)^8} h_2(x) + \frac{33}{140(x^2)^9} h_3(x) \right\}. \tag{3.12}$$

The subtracted sum

$$A_0' = \sideset{}{'}\sum_{x \in \Lambda} \left[ G(x)^3 - \{G(x)^3\}_{\text{as}} \right] \tag{3.13}$$

is hence rapidly convergent and so may be computed accurately by summing over all points $x$ with $|x| \leq 20$. The subtraction term

$$A_0'' = {\sum_{x \in \Lambda}}' \{G(x)^3\}_{\text{as}} \tag{3.14}$$

may be expressed as a sum of zeta functions,

$$A_0'' = \frac{1}{(4\pi^2)^3} \Big\{ Z(3, h_0) + \frac{3}{10} Z(5, h_0) + 3 Z(6, h_1)$$

$$+ \frac{24}{7} Z(7, h_1) - \frac{3}{4} Z(8, h_2) + \frac{33}{140} Z(9, h_3) \Big\}, \tag{3.15}$$

and is thus accurately calculable, too. Noting

$$A_0 = A_0' + A_0'' + G(0)^3, \tag{3.16}$$

the final result then is

$$A_0 = 0.0040430548122(3). \tag{3.17}$$

A more precise determination of $A_0$ along these lines would of course be possible, but it seems unlikely that this will ever be required.

### 3.3 More complicated integrals

The techniques which have been discussed so far can easily be generalized to compute integrals of the form

$$\int_{-\pi}^{\pi} \frac{\mathrm{d}^4 k}{(2\pi)^4} \frac{\mathrm{d}^4 q}{(2\pi)^4} \frac{P(k, q, r)}{(\hat{k}^2)^n (\hat{q}^2)^m (\hat{r}^2)^l}, \tag{3.18}$$

where $P(k, q, r)$ is a polynomial in the sines and cosines of the momenta $k, q, r$ and $n, m, l$ are positive integers ($r = -k - q$ as before). To illustrate this we consider the integrals

$$Q_1 = \int_{-\pi}^{\pi} \frac{\mathrm{d}^4 k}{(2\pi)^4} \frac{\mathrm{d}^4 q}{(2\pi)^4} \sum_{\mu=0}^{3} \frac{\hat{k}_\mu^2 \hat{q}_\mu^2}{\hat{k}^2 \hat{q}^2 \hat{r}^2}, \tag{3.19}$$

$$Q_2 = \int_{-\pi}^{\pi} \frac{\mathrm{d}^4 k}{(2\pi)^4} \frac{\mathrm{d}^4 q}{(2\pi)^4} \sum_{\mu=0}^{3} \frac{\hat{k}_\mu^2 \hat{q}_\mu^2 \hat{r}_\mu^2}{\hat{k}^2 \hat{q}^2 \hat{r}^2}, \qquad (3.20)$$

$$R_1 = \int_{-\pi}^{\pi} \frac{\mathrm{d}^4 k}{(2\pi)^4} \frac{\mathrm{d}^4 q}{(2\pi)^4} \sum_{\mu=0}^{3} \frac{\hat{k}_\mu^2 \hat{q}_\mu^2}{(\hat{k}^2)^2 \hat{q}^2 \hat{r}^2}, \qquad (3.21)$$

$$T_3 = \int_{-\pi}^{\pi} \frac{\mathrm{d}^4 k}{(2\pi)^4} \frac{\mathrm{d}^4 q}{(2\pi)^4} \sum_{\mu=0}^{3} \frac{\hat{k}_\mu^4}{(\hat{k}^2)^3 \hat{q}^2 \hat{r}^2}. \qquad (3.22)$$

In position space the first of these is given by

$$Q_1 = \sum_{x \in \Lambda} \sum_{\mu=0}^{3} \left[-\nabla_\mu^* \nabla_\mu G(x)\right]^2 G(x) \qquad (3.23)$$

[cf. eq.(2.1)]. The terms in this sum fall off like $|x|^{-10}$ and a direct evaluation without subtraction yields

$$Q_1 = 0.0423063684(1). \qquad (3.24)$$

In the same way the result

$$Q_2 = 0.054623978180(1) \qquad (3.25)$$

is obtained.

To compute the integral (3.21) we introduce the auxiliary function

$$K(x) = \int_{-\pi}^{\pi} \frac{\mathrm{d}^4 p}{(2\pi)^4} \left(\mathrm{e}^{ipx} - 1\right) / (\hat{p}^2)^2. \qquad (3.26)$$

The position space representation of $R_1$ then reads

$$R_1 = \sum_{x \in \Lambda} \sum_{\mu=0}^{3} \left[-\nabla_\mu^* \nabla_\mu K(x)\right] \left[-\nabla_\mu^* \nabla_\mu G(x)\right] G(x). \qquad (3.27)$$

Some properties of the function $K(x)$ are discussed in appendix C. In particular, it is shown there that $K$ is recursively computable in terms of the propagator $G$. We may thus apply the techniques explained above and obtain

$$R_1 = 0.006603075727(1), \qquad (3.28)$$

where a subtraction has been made to accelerate the convergence of the sum (3.27).

In the case of the integral (3.22) another auxiliary function,

$$L(x) = \int_{-\pi}^{\pi} \frac{d^4 p}{(2\pi)^4} \left( e^{ipx} - 1 - i\mathring{p}x + \tfrac{1}{2}(\mathring{p}x)^2 \right) / (\hat{p}^2)^3, \tag{3.29}$$

must be introduced. The notation

$$\mathring{p}_\mu = \sin(p_\mu), \qquad \mathring{p}x = \sum_{\mu=0}^{3} \mathring{p}_\mu x_\mu, \tag{3.30}$$

has here been used. This function is also calculable in much the same way as $K$ (appendix C). In position space we have

$$T_3 = \sum_{x \in \Lambda} \sum_{\mu=0}^{3} [\nabla_\mu^* \nabla_\mu \nabla_\mu^* \nabla_\mu L(x)] G(x)^2, \tag{3.31}$$

and the result

$$T_3 = 0.00173459425(1) \tag{3.32}$$

is now obtained straightforwardly by applying the subtraction method.

## 4. Feynman integrals with non-zero external momenta

In this section we consider the Feynman diagrams shown in fig. 1. The external momentum is denoted by $p$ and the numerators of the corresponding Feynman integrands are set equal to 1. We are interested in the continuum limit of these integrals. Since we are using lattice units and since there are no mass parameters, the continuum limit is equivalent to taking $p$ to zero, i.e. our task is to work out the leading terms in an asymptotic expansion of the diagrams for $p \to 0$.

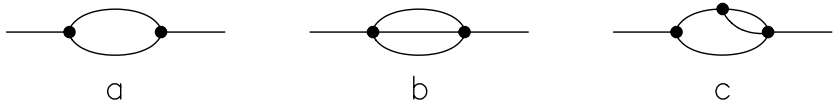

Fig. 1. Self-energy diagrams studied in sect. 4. (b) and (c) are referred to as the "bigmac" and the "eye" diagram, respectively.

### 4.1. Scalar one-loop integral

The integral corresponding to the first diagram in fig. 1 is given by

$$A(p) = \int_{-\pi}^{\pi} \frac{d^4k}{(2\pi)^4} \frac{1}{\hat{k}^2 \hat{s}^2}, \qquad s = k + p. \tag{4.1}$$

$A(p)$ is a smooth function at all momenta $p$ that do not vanish modulo $2\pi$. We would like to show that

$$A(p) \underset{p \to 0}{=} \frac{1}{(4\pi)^2} \left[ -\ln(p^2) + 2 \right] + P_2 + O(p^2), \tag{4.2}$$

where $P_2$ is a constant [in this section $O(p^n)$ stands for a remainder $R(p)$ such that $\lim_{p \to 0} R(p)/|p|^{n-\epsilon} = 0$ for all $\epsilon > 0$].

We first compute $A(p)$ in a conventional way using the Reisz power counting theorem [3,4]. The idea is to subtract the integral at $p = 0$ to make it convergent in the continuum limit. We thus define a subtracted integral,

$$A'(p, m) = \int_{-\pi}^{\pi} \frac{d^4k}{(2\pi)^4} \left\{ \frac{1}{\hat{k}^2 \hat{s}^2} - \frac{1}{(\hat{k}^2 + m^2)^2} \right\}, \tag{4.3}$$

and the subtraction term

$$A''(m) = \int_{-\pi}^{\pi} \frac{d^4k}{(2\pi)^4} \frac{1}{(\hat{k}^2 + m^2)^2}, \tag{4.4}$$

where $m > 0$ is an auxiliary mass parameter. The full integral

$$A(p) = A'(p, m) + A''(m) \tag{4.5}$$

is of course independent of $m$. We are thus free scale $m$ to zero proportionally to $p$. In this limit the Reisz power counting theorem applies and one concludes

that

$$\lim_{\lambda \to 0} A'(\lambda p, \lambda m) = \int_{-\infty}^{\infty} \frac{d^4 k}{(2\pi)^4} \left\{ \frac{1}{k^2 s^2} - \frac{1}{(k^2 + m^2)^2} \right\}$$

$$= \frac{1}{(4\pi)^2} \left[ \ln(m^2/p^2) + 2 \right]. \tag{4.6}$$

The subtraction term, on the other hand, diverges logarithmically for $m \to 0$. As shown in appendix D the divergence is such that the limit

$$P_2 = \lim_{m \to 0} \left\{ \frac{1}{(4\pi)^2} \ln(m^2) + A''(m) \right\} \tag{4.7}$$

exists. Numerically one obtains

$$P_2 = 0.02401318111946489(1). \tag{4.8}$$

Putting these results together, we conclude that eq.(4.2) holds except perhaps for the form of the error term (at this point we only know it vanishes for $p \to 0$).

Eq.(4.2) may also easily be derived using coordinate space methods. The starting point is the representation

$$A(p) = \lim_{\epsilon \to 0} \sum_{x \in \Lambda} e^{-\epsilon x^2} e^{-ipx} G(x)^2. \tag{4.9}$$

We then note that the function $H(x)$ [eq.(2.7)] satisfies

$$H(x) \underset{x \to \infty}{\sim} -\frac{1}{\pi^2 (x^2)^2} \left\{ 1 - \frac{4}{x^2} + 8 \frac{x^4}{(x^2)^3} \right.$$

$$\left. - \frac{7}{(x^2)^2} + 40 \frac{x^4}{(x^2)^4} - 288 \frac{x^6}{(x^2)^5} + 280 \frac{(x^4)^2}{(x^2)^6} + \ldots \right\} \tag{4.10}$$

and so has the same leading order behaviour as $G(x)^2$. The convergence of the sum (4.9) can thus be accelerated by separating a singular part $A_s(p)$ according to

$$A(p) = A_s(p) + A_r(p), \tag{4.11}$$

where

$$A_s(p) = -(4\pi)^{-2} \lim_{\epsilon \to 0} \sum_{x \in \Lambda} e^{-\epsilon x^2} e^{-ipx} H(x), \qquad (4.12)$$

$$A_r(p) = \sum_{x \in \Lambda} e^{-ipx} \left[ G(x)^2 + (4\pi)^{-2} H(x) \right]. \qquad (4.13)$$

Since $A_s(p)$ is just the Fourier transform of $H(x)$ and since the square bracket in the last expression above falls off like $|x|^{-6}$, we deduce that

$$A(p) \underset{p \to 0}{=} -\frac{1}{(4\pi)^2} \ln(p^2) + A_r(0) + \mathrm{O}(p^2). \qquad (4.14)$$

We have thus rederived eq.(4.2) with the constant $P_2$ being given by

$$P_2 = -\frac{2}{(4\pi)^2} + \sum_{x \in \Lambda} \left[ G(x)^2 + (4\pi)^{-2} H(x) \right]. \qquad (4.15)$$

Moreover the argument shows that the error term is as quoted. The next-to-leading terms in the expansion could in fact easily be calculated by making a further subtraction. We finally remark that the sum in eq.(4.15) can be computed using the techniques of sect. 3.

*4.2 The bigmac diagram*

We now proceed to discuss the integral

$$D(p) = \int_{-\pi}^{\pi} \frac{\mathrm{d}^4 k}{(2\pi)^4} \frac{\mathrm{d}^4 q}{(2\pi)^4} \frac{1}{\hat{k}^2 \hat{q}^2 \hat{r}^2}, \qquad r = -k - q - p \qquad (4.16)$$

(cf. fig. 1b). Our goal is to establish the expansion

$$D(p) \underset{p \to 0}{=} A_0 + p^2 \left[ A_1 + B_1 \ln(p^2) \right] + \mathrm{O}(p^4) \qquad (4.17)$$

and to compute the coefficients $A_1$ and $B_1$ (the leading term $A_0$ has already been calculated in sect. 3).

The situation is very much the same as in the case of the one-loop integral $A(p)$ discussed above. We first note that

$$D(p) = \sum_{x \in \Lambda} e^{-ipx} G(x)^3 \qquad (4.18)$$

and split the sum into a singular and a regular part according to

$$D(p) = D_s(p) + D_r(p), \qquad (4.19)$$

where

$$D_s(p) = -\tfrac{1}{2}(4\pi)^{-4} \sum_{x \in \Lambda} e^{-ipx} \Delta H(x), \qquad (4.20)$$

$$D_r(p) = \sum_{x \in \Lambda} e^{-ipx} \left[ G(x)^3 + \tfrac{1}{2}(4\pi)^{-4} \Delta H(x) \right]. \qquad (4.21)$$

The splitting has been done in such a way that the square bracket in the last expression falls off like $|x|^{-8}$ at large $x$ [cf. eqs.(3.12),(4.10)]. The expansion of $D_r(p)$ to order $p^2$ is thus obtained straightforwardly by expanding the exponential factor in eq.(4.21). As for the singular part we note that $D_s(p)$ is just the Fourier transform of $\Delta H(x)$, i.e. $D_s(p)$ is proportional to $\hat{p}^2 \ln(\hat{p}^2)$.

Taken together our discussion shows that the expansion (4.17) holds with

$$A_1 = -\tfrac{1}{8} \sum_{x \in \Lambda} x^2 \left[ G(x)^3 + \tfrac{1}{2}(4\pi)^{-4} \Delta H(x) \right] \qquad (4.22)$$

and $B_1 = \tfrac{1}{2}(4\pi)^{-4}$. To calculate the sum (4.22) we again make use of the techniques introduced in sect. 3. Note that $H(x)$ is computable at all non-zero $x$ through eq.(2.6). $H(0)$ is calculated in appendix D. In this way the result

$$A_1 = -0.00007447695(1) \qquad (4.23)$$

is obtained.

*4.3 The eye diagram*

The integral corresponding to the scalar eye diagram (fig. 1c) may be written in the form

$$E(p) = \int_{-\pi}^{\pi} \frac{d^4 k}{(2\pi)^4} \frac{1}{\hat{k}^2 \hat{s}^2} A(k), \qquad (4.24)$$

where $A(k)$ denotes the one-loop integral studied previously and $s = k + p$ as before. In the following our aim is to show that

$$E(p) \underset{p \to 0}{=} \frac{1}{2(4\pi)^4} \left[\ln(p^2)\right]^2 - \frac{1}{(4\pi)^2} \left[ P_2 + \frac{3}{(4\pi)^2} \right] \ln(p^2) + X_1 + \mathrm{O}(p^2) \quad (4.25)$$

and to calculate the constant $X_1$.

We begin by defining the function

$$A_{\text{as}}(k) = \frac{1}{(4\pi)^2}\left[-\ln(\hat{k}^2) + 2\right] + P_2. \tag{4.26}$$

As discussed in subsect. 4.1 the one-loop integral $A(k)$ satisfies

$$A(k) - A_{\text{as}}(k) \underset{k\to 0}{=} \text{O}(k^2). \tag{4.27}$$

This suggests that we introduce singular and regular parts of the eye diagram through

$$E_s(p) = -\frac{1}{(4\pi)^2}\int_{-\pi}^{\pi}\frac{d^4k}{(2\pi)^4}\frac{\ln(\hat{k}^2)}{\hat{k}^2\,\hat{s}^2}, \tag{4.28}$$

$$E_r(p) = \int_{-\pi}^{\pi}\frac{d^4k}{(2\pi)^4}\frac{1}{\hat{k}^2\,\hat{s}^2}\left[A(k) - A_{\text{as}}(k)\right]. \tag{4.29}$$

The relation

$$E(p) = E_s(p) + E_r(p) + \left[P_2 + \frac{2}{(4\pi)^2}\right]A(p) \tag{4.30}$$

is an immediate consequence of these definitions.

We first consider the regular part $E_r(p)$. The function $A(k) - A_{\text{as}}(k)$ is smooth in the whole integration range except at $k = 0$ where its behaviour is given by eq.(4.27). Taking this into account it is easy to show that

$$E_r(p) \underset{p\to 0}{=} E_r(0) + \text{O}(p^2). \tag{4.31}$$

The constant $E_r(0)$ can be computed using position space techniques. From the discussion in subsect. 4.1 we infer that

$$A(k) - A_{\text{as}}(k) = \sum_{x\in\Lambda}\left(e^{-ikx} - 1\right)\left[G(x)^2 + (4\pi)^{-2}H(x)\right]. \tag{4.32}$$

As already noted before the square bracket in this expression falls off like $|x|^{-6}$ at large $x$. When eq.(4.32) is inserted in eq.(4.29), the integration and summation may hence be interchanged and it follows that

$$E_r(0) = \sum_{x\in\Lambda}K(x)\left[G(x)^2 + (4\pi)^{-2}H(x)\right]. \tag{4.33}$$

The auxiliary function $K(x)$ has been introduced in subsect. 3.3 and is discussed further in appendix C. At large $x$ the terms in the sum (4.33) have an asymptotic expansion consisting of two series of the usual form, of which one is multiplied by $\ln(x^2)$ [cf. eq.(C.8)]. The subtraction method of sect. 3 may be applied in such cases as well. We only have to introduce the derivatives of the appropriate zeta functions to calculate the subtraction term. The result

$$E_r(0) = -0.000002546129(1) \qquad (4.34)$$

is thus obtained.

We still need to evaluate the singular part $E_s(p)$. This is a one-loop integral which may be calculated using the momentum space subtraction technique previously employed in the case of the integral $A(p)$ (subsect. 4.1). We thus split the integral into three parts,

$$E_s(p) = -\frac{1}{(4\pi)^2} \left\{ E'_s(p,m) + E''_s(m) + \ln(m^2) A(p) \right\}, \qquad (4.35)$$

where

$$E'_s(p,m) = \int_{-\pi}^{\pi} \frac{d^4k}{(2\pi)^4} \left\{ \frac{\ln(\hat{k}^2/m^2)}{\hat{k}^2 \hat{s}^2} - \frac{\ln(\hat{k}^2/m^2+1)}{(\hat{k}^2+m^2)^2} \right\}, \qquad (4.36)$$

$$E''_s(m) = \int_{-\pi}^{\pi} \frac{d^4k}{(2\pi)^4} \frac{\ln(\hat{k}^2/m^2+1)}{(\hat{k}^2+m^2)^2}. \qquad (4.37)$$

The auxiliary mass parameter $m > 0$ is again scaled to zero in a fixed proportion to $p$. From the Reisz power counting theorem we then infer that

$$\lim_{\lambda \to 0} E'_s(\lambda p, \lambda m) = \int_{-\infty}^{\infty} \frac{d^4k}{(2\pi)^4} \left\{ \frac{\ln(k^2/m^2)}{k^2 s^2} - \frac{\ln(k^2/m^2+1)}{(k^2+m^2)^2} \right\}$$

$$= -\frac{1}{(4\pi)^2} \left\{ \tfrac{1}{2} \left[\ln(m^2/p^2)\right]^2 + \ln(m^2/p^2) \right\}. \qquad (4.38)$$

The second part, $E''_s(m)$, is logarithmically divergent for $m \to 0$ in such a way that the limit

$$P_3 = \lim_{m \to 0} \left\{ -\frac{1}{2(4\pi)^2} \left[\ln(m^2)\right]^2 + \left[P_2 + \frac{1}{(4\pi)^2}\right] \ln(m^2) + E''_s(m) \right\} \qquad (4.39)$$

exists. This is shown in appendix D where $P_3$ is also computed numerically. Collecting all contributions to eq.(4.35) the result

$$E_s(p) \underset{p \to 0}{=} \frac{1}{(4\pi)^4} \left\{ \tfrac{1}{2} \left[\ln(p^2)\right]^2 - \ln(p^2) \right\} - \frac{1}{(4\pi)^2} P_3 + \mathrm{O}(p^2) \qquad (4.40)$$

is thus obtained.

We finally go back to eq.(4.30) and conclude that the expansion (4.25) holds with

$$X_1 = E_r(0) + \left[P_2 + \frac{2}{(4\pi)^2}\right]^2 - \frac{1}{(4\pi)^2} P_3. \qquad (4.41)$$

Numerically one obtains

$$X_1 = 0.001123187953(1). \qquad (4.42)$$

As a last point we mention that $X_1$ has previously been computed by Allés et al. [5] through numerical integration in momentum space. Their result was estimated to be accurate to 4 significant decimal places but differs from the number given above by 10%. We do not know the cause for this discrepancy.

## 5. Concluding remarks

In this paper attention has been restricted to lattice theories in four dimensions with free propagators given by eq.(1.3). The extension of our methods to other situations is however straightforward. In particular, the recursive computation of free propagators, such as the Dirac propagator in lattice QCD with Wilson quarks, is generally possible along the lines discussed in sect. 2. An amusing remark in this connection is that in two dimensions the propagator of a scalar massless field with standard lattice action is at any point $x$ exactly computable as a linear combination of $1/\pi$ and 1 with rational coefficients.

Most diagrams that we encountered in our computation of the relation between the bare lattice coupling and the $\overline{\mathrm{MS}}$ coupling in $\mathrm{SU}(N)$ gauge theories [1] have integrands with complicated numerators. It turns out, however, that the techniques introduced in this paper are sufficient to compute all of them. The important point to note is that the continuum limit of most of these integrals is rather easy to take. The resulting expansion coefficients are integrals of the type discussed in sect. 3 and so can be calculated by the methods described there. Further details on these computations will be published elsewhere.

We are indebted to Claus Vohwinkel for sharing his insights about the free propagator with us.

## Appendix A

In this appendix we derive the large $x$ expansion of the free propagator given in eq.(2.4). The methods employed are more generally useful. We begin with the expansion of some auxiliary functions and then consider the propagator.

*A.1 Auxiliary functions*

Let $Q(p)$ be a homogeneous polynomial in the momentum $p$ with degree $d$, and $n \geq 1$ an integer such that $d - 2n \geq -3$. The integral

$$I(x) = \int_{-\infty}^{\infty} \frac{\mathrm{d}^4 p}{(2\pi)^4} \mathrm{e}^{ipx} \mathrm{e}^{-p^2} Q(p)(p^2)^{-n} \tag{A.1}$$

is then absolutely convergent and well-defined for all $x \in \mathbb{R}^4$. Our aim in the following is to establish the behaviour of $I(x)$ for large $x$.

Consider first the case $n = 1$. Here we have

$$\begin{aligned} I(x) &= Q(-i\partial) \int_1^{\infty} \mathrm{d}t \int_{-\infty}^{\infty} \frac{\mathrm{d}^4 p}{(2\pi)^4} \mathrm{e}^{ipx} \mathrm{e}^{-tp^2} \\ &= Q(-i\partial)(4\pi^2 x^2)^{-1} \left(1 - \mathrm{e}^{-x^2/4}\right), \end{aligned} \tag{A.2}$$

from which it follows that

$$I(x) \underset{x \to \infty}{\sim} Q(-i\partial)(4\pi^2 x^2)^{-1} \tag{A.3}$$

up to exponentially small corrections.

Let us now assume that $n \geq 2$. Define $m = 2n - 4$ and let $T_m$ be the Taylor operator of order $m$, viz.

$$T_m f(x) = \sum_{k=0}^{m} \frac{1}{k!} x_{\mu_1} \ldots x_{\mu_k} \partial_{\mu_1} \ldots \partial_{\mu_k} f(0). \qquad (A.4)$$

Since $d > m$ we have

$$Q(p) e^{ipx} = Q(-i\partial)(1 - T_m) e^{ipx} \qquad (A.5)$$

and so conclude that

$$I(x) = Q(-i\partial) \frac{1}{(n-1)!} \int_1^\infty dt\, (t-1)^{n-1} (4\pi t)^{-2} (1 - T_m) e^{-x^2/4t}. \qquad (A.6)$$

Note that the differentiation and integration could be interchanged here because $(1 - T_m) e^{ipx} e^{-p^2} (p^2)^{-n}$ is absolutely integrable. Next we substitute $t \to 1/t$ and use the binomial formula. This yields

$$I(x) = \frac{1}{(4\pi)^2} \sum_{\mu=0}^{n-1} \frac{(-1)^{n-1-\mu}}{\mu!(n-1-\mu)!} Q(-i\partial)$$

$$\times \int_0^1 dt\, t^{-\mu} \left\{ e^{-tx^2/4} - \sum_{\nu=0}^{\mu-1} \frac{1}{\nu!} (-tx^2/4)^\nu \right\}. \qquad (A.7)$$

Through partial integration all terms with $\mu \geq 1$ can be reduced to

$$\int_0^1 dt\, t^{-1} \left( e^{-tx^2/4} - 1 \right) \underset{x \to \infty}{\sim} -\ln(x^2) + \text{constant} + \ldots$$

In this way we obtain the asymptotic expansion

$$I(x) \underset{x \to \infty}{\sim} \frac{1}{(4\pi)^2} Q(-i\partial) \Biggl\{ \frac{(-1)^{n-1}}{(n-1)!} \cdot \frac{4}{x^2}$$

$$+ \sum_{\mu=1}^{n-1} \frac{(-1)^{n-1}}{(\mu-1)!\mu!(n-1-\mu)!} \left( \frac{x^2}{4} \right)^{\mu-1} \ln(x^2) \Biggr\}, \qquad (A.8)$$

which is valid up to exponentially small corrections. Note that after differentiation, the logarithm disappears and one is left with terms of the form $P(x)(x^2)^{-k}$, where $P$ is a polynomial and $k$ an integer.

*A.2 Large $x$ expansion of the propagator*

Let $h(p)$, $p \in \mathbb{R}^4$, be a smooth function with

$$h(p) = \begin{cases} 1 & \text{if } |p| < 1, \\ 0 & \text{if } |p| > 2. \end{cases} \qquad (A.9)$$

$G(x)$ then coincides with

$$G_h(x) = \int_{-\infty}^{\infty} \frac{d^4p}{(2\pi)^4} e^{ipx} h(p)(\hat{p}^2)^{-1} \qquad (A.10)$$

to all orders of $|x|^{-1}$. Next we note that

$$\hat{p}^2 = \sum_{n=1}^{\infty} (-1)^{n+1} \frac{2}{(2n)!} p^{2n}, \qquad (A.11)$$

where we have introduced the notation

$$p^n = \sum_{\mu=0}^{3} (p_\mu)^n. \qquad (A.12)$$

It follows that

$$e^{p^2}(\hat{p}^2)^{-1} \underset{p \to 0}{=} (p^2)^{-1} + 1 + \frac{1}{12}p^4(p^2)^{-2} + \frac{1}{2}p^2 + \frac{1}{12}p^4(p^2)^{-1}$$

$$- \frac{1}{360}p^6(p^2)^{-2} + \frac{1}{144}(p^4)^2(p^2)^{-3} + \ldots \qquad (A.13)$$

When this series is multiplied with $e^{-p^2}$ and subtracted from the integrand in eq.(A.10), the singularity at $p = 0$ is smoothed. As a result we have

$$G(x) \underset{x \to \infty}{\sim} \int_{-\infty}^{\infty} \frac{d^4p}{(2\pi)^4} e^{ipx} e^{-p^2} \left\{ \frac{1}{p^2} + \frac{1}{12} \frac{p^4}{(p^2)^2} + \ldots \right\} \qquad (A.14)$$

and hence, using the results of subsection A.1,

$$G(x) \underset{x \to \infty}{=} \frac{1}{4\pi^2 x^2} \left\{ 1 - \frac{x^2}{48} \partial^4 \ln(x^2) \right.$$

$$\left. - \frac{x^2}{1440} \partial^6 \ln(x^2) + \frac{x^2}{4608}(\partial^4)^2 x^2 \ln(x^2) \right\} + O(|x|^{-8}). \qquad (A.15)$$

Carrying out the differentiations one obtains the expansion (2.4). Higher order terms could be generated easily, using a symbolic manipulation program.

### Appendix B

In this appendix we establish the relations (2.12) and (2.13) following an argument first given by Vohwinkel [2]. The starting point is the observation that the recursion relation (2.9) becomes one-dimensional along the lattice axes. Explicitly, if we define

$$g_0(n) = G(n, 0, 0, 0), \tag{B.1}$$

$$g_1(n) = G(n, 1, 0, 0), \tag{B.2}$$

$$g_2(n) = G(n, 1, 1, 0), \tag{B.3}$$

$$g_3(n) = G(n, 1, 1, 1), \tag{B.4}$$

and use the lattice symmetries, it is easy to show that

$$g_0(n+1) = 8g_0(n) - 6g_1(n) - g_0(n-1), \tag{B.5}$$

$$g_1(n+1) = \frac{2n}{n+1}[4g_1(n) - g_0(n) - 2g_2(n)] - \frac{n-1}{n+1}g_1(n-1), \tag{B.6}$$

$$g_2(n+1) = \frac{2n}{n+2}[4g_2(n) - 2g_1(n) - g_3(n)] - \frac{n-2}{n+2}g_2(n-1), \tag{B.7}$$

$$g_3(n+1) = \frac{2n}{n+3}[4g_3(n) - 3g_2(n)] - \frac{n-3}{n+3}g_3(n-1), \tag{B.8}$$

for all $n \geq 1$. These equations look deceptively simple, but so far we have not been able to solve them exactly. Vohwinkel has however noticed that

$$\begin{aligned} k_1(n) = & (n-1)g_0(n) + 3ng_1(n) + 3(n+1)g_2(n) + (n+2)g_3(n) \\ & - ng_0(n-1) - 3(n-1)g_1(n-1) \\ & - 3(n-2)g_2(n-1) - (n-3)g_3(n-1) \end{aligned} \tag{B.9}$$

is a "constant of motion", i.e. using the recursion relations one may show that

$$k_1(n+1) = k_1(n) \quad \text{for} \quad n \geq 1. \tag{B.10}$$

Another such constant of motion is given by

$$k_2(n) = n^2 k_1(n) - 2(n-1)(n-2)g_0(n) - 3n(2n-3)g_1(n)$$
$$- 6(n+1)(n-1)g_2(n) - (n+2)(2n-1)g_3(n) - 3ng_0(n-1)$$
$$- 6(n-1)g_1(n-1) - 3(n-2)g_2(n-1), \tag{B.11}$$

but no further independent constant was found.

The values of $k_1$ and $k_2$ are easily worked out in the limit $n \to \infty$ where

$$g_\mu(n) = (4\pi^2 n^2)^{-1} + \mathrm{O}(1/n^3) \quad \text{for all } \mu = 0, \ldots, 3. \tag{B.12}$$

The result is

$$k_1 = 0, \tag{B.13}$$

$$k_2 = -4/\pi^2, \tag{B.14}$$

and setting $n = 1$ in eqs.(B.9) and (B.11) now directly yields eqs.(2.12) and (2.13).

## Appendix C

*C.1 Properties of $K(x)$*

From the definition (3.26) of the auxiliary function $K(x)$ it is straightforward to show that

$$-\Delta K(x) = G(x) \tag{C.1}$$

and

$$(\nabla_\mu^* + \nabla_\mu)K(x) = -x_\mu G(x). \tag{C.2}$$

Using the latter equation we can determine $K(x)$ at any point $x$ recursively in terms of $G$ and the values of $K$ at the corners of the unit hypercube at the origin. These are given by

$$K(0,0,0,0) = 0, \tag{C.3}$$

$$K(1,0,0,0) = -\frac{1}{8}G(0,0,0,0), \tag{C.4}$$

$$K(1,1,0,0) = -\frac{1}{6}G(0,0,0,0), \tag{C.5}$$

$$K(1,1,1,0) = -\frac{5}{24}G(0,0,0,0) + \frac{1}{4}G(1,1,0,0), \tag{C.6}$$

$$K(1,1,1,1) = \frac{1}{6}G(0,0,0,0) - \frac{1}{2}G(1,1,0,0) - \frac{1}{2\pi^2}. \tag{C.7}$$

$K(x)$ is thus obtained as a linear combination of the form (2.14) and so is accurately computable.

At large $x$ the asymptotic expansion

$$K(x) \underset{x \to \infty}{=} -\frac{1}{(4\pi)^2} \left\{ \ln(x^2) + c_1 - \frac{2x^4}{3(x^2)^3} + \mathrm{O}(|x|^{-4}) \right\} \tag{C.8}$$

holds, where

$$c_1 = 3.560146537656400(2). \tag{C.9}$$

Eq.(C.8) has been derived by applying the techniques discussed in appendix A. The leading singularity at $p = 0$ in the defining integral (3.26) (which gives rise to the logarithm in the large $x$ expansion) was taken care of by introducing an appropriate auxiliary function $I(x)$.

C.2 Properties of $L(x)$

The function $L(x)$ defined through eq.(3.29) satisfies

$$-\Delta L(x) = K(x) + \frac{1}{8}G(0) - \frac{1}{32\pi^2} \tag{C.10}$$

and

$$(\nabla^*_\mu + \nabla_\mu)L(x) = -\frac{1}{2}x_\mu \left[ K(x) + \frac{1}{8}G(0) \right]. \tag{C.11}$$

Its values at the corners of the unit hypercube at the origin are

$$L(0,0,0,0) = 0, \tag{C.12}$$

$$L(1,0,0,0) = -\frac{1}{64}G(0,0,0,0) + \frac{1}{256\pi^2}, \tag{C.13}$$

$$L(1,1,0,0) = -\frac{1}{48}G(0,0,0,0) + \frac{1}{96\pi^2}, \tag{C.14}$$

$$L(1,1,1,0) = -\frac{5}{192}G(0,0,0,0) + \frac{19}{768\pi^2}, \tag{C.15}$$

$$L(1,1,1,1) = -\frac{1}{16}G(0,0,0,0) + \frac{1}{16}G(1,1,0,0) + \frac{1}{12\pi^2}. \tag{C.16}$$

This function is, therefore, also expressible as a linear combination of the form (2.14).

The asymptotic behaviour of $L(x)$ is given by

$$L(x) \underset{x\to\infty}{=} \frac{1}{128\pi^2} \left\{ x^2 \ln(x^2) + c_2 x^2 - \ln(x^2) + c_3 + \frac{x^4}{3(x^2)^2} + \mathrm{O}(|x|^{-2}) \right\} \tag{C.17}$$

with

$$c_2 = -0.49811600254393(2), \tag{C.18}$$

$$c_3 = -2.77499211742106(1). \tag{C.19}$$

### Appendix D

The computation of the constants $P_2$, $P_3$ and $H(0)$ discussed in this appendix is based on the identity

$$\int_{-\pi}^{\pi} \frac{\mathrm{d}^4 k}{(2\pi)^4}(\hat{k}^2 + m^2)^{-\alpha} = \frac{1}{\Gamma(\alpha)} \int_0^\infty \mathrm{d}t\, t^{\alpha-1} \mathrm{e}^{-tm^2} \left[ \mathrm{e}^{-2t} I_0(2t) \right]^4, \tag{D.1}$$

which holds for all $\alpha > 0$ and $m > 0$. The function $I_0(z)$ appearing in this formula is a modified Bessel function (ref.[6], §8.43). To evaluate the constant

$$P_2 = \lim_{m \to 0} \left\{ \frac{1}{(4\pi)^2} \ln(m^2) + \int_{-\pi}^{\pi} \frac{d^4k}{(2\pi)^4} \frac{1}{(\hat{k}^2 + m^2)^2} \right\}, \tag{D.2}$$

we insert eq.(D.1) with $\alpha = 2$ and discuss the limiting behaviour of the integral for $m \to 0$. The Bessel function $I_0(z)$ is a smooth function with asymptotic behaviour such that

$$\left[ e^{-2t} I_0(2t) \right]^4 \underset{t \to \infty}{=} (4\pi t)^{-2} + O(t^{-3}). \tag{D.3}$$

It follows from this that the integral diverges as $m \to 0$, but it is now easy to show that the limit (D.2) exists and is given by

$$P_2 = -\frac{\gamma}{(4\pi)^2} + \int_0^\infty dt\, t \left\{ \left[ e^{-2t} I_0(2t) \right]^4 - \theta(t-1)(4\pi t)^{-2} \right\}, \tag{D.4}$$

where $\gamma = -\Gamma'(1)$ denotes Euler's constant. From this representation it is straightforward to compute $P_2$ by numerical integration, using library routines for the Bessel function. As a result one obtains the number quoted in eq.(4.8).

In the case of the constant $P_3$ [eq.(4.39)] we differentiate the identity (D.1) with respect to $\alpha$ at $\alpha = 2$ and proceed as above. In a few steps, taking eq.(D.4) into account, one obtains

$$P_3 = \frac{1}{(4\pi)^2} \left[ \frac{1}{2}\gamma^2 - \gamma - \frac{\pi^2}{12} \right]$$

$$+ \int_0^\infty dt\, t[1 - \gamma - \ln(t)] \left\{ \left[ e^{-2t} I_0(2t) \right]^4 - \theta(t-1)(4\pi t)^{-2} \right\}, \tag{D.5}$$

and numerical integration now yields

$$P_3 = 0.03467243216643385(1) \tag{D.6}$$

(for better convergence the series expansions of the Bessel function have been used in the range $0 \leq t \leq 0.01$ and $100 \leq t < \infty$).

We finally discuss the computation of the constant

$$H(0) = \int_{-\pi}^{\pi} \frac{d^4k}{(2\pi)^4} \ln(\hat{k}^2). \tag{D.7}$$

In this case we first restrict the parameter $\alpha$ in eq.(D.1) to the range $0 < \alpha < 1$. We can then set $m = 0$ in the formula. After performing a subtraction at $t = 0$ the identity becomes

$$\int_{-\pi}^{\pi} \frac{\mathrm{d}^4 k}{(2\pi)^4} (\hat{k}^2)^{-\alpha} = \frac{1}{\Gamma(\alpha + 1)}$$
$$+ \frac{1}{\Gamma(\alpha)} \int_0^\infty \mathrm{d}t \, t^{\alpha-1} \left\{ \left[ \mathrm{e}^{-2t} I_0(2t) \right]^4 - \theta(1-t) \right\}, \qquad \text{(D.8)}$$

which is now valid for $-1 < \alpha < 1$. Differenting at $\alpha = 0$ we obtain

$$H(0) = -\gamma - \int_0^\infty \mathrm{d}t \, t^{-1} \left\{ \left[ \mathrm{e}^{-2t} I_0(2t) \right]^4 - \theta(1-t) \right\} \qquad \text{(D.9)}$$

and numerical integration gives

$$H(0) = 1.99970764451731256(1). \qquad \text{(D.10)}$$